# Characterizing the Chemistry of Planetary Materials Around White Dwarf Stars

B. Zuckerman[1] and E.D. Young[2]

[1]Department of Physics & Astronomy, University of California, Los Angeles, Los Angeles, CA 90095, USA

[2] Department of Earth, Planetary, and Space Sciences, University of California, Los Angeles, Los Angeles, CA 90095, USA

**Abstract**

Planetary systems that orbit white dwarf stars can be studied via spectroscopic observations of the stars themselves. Numerous white dwarfs are seen to have accreted mostly rocky minor planets, the remnants of which are present in the stellar photospheres. The elemental abundances in the photospheres unveil the bulk compositions of the accreted parent bodies with a precision far greater than can be attained with any other technique currently available to astronomers. The most significant discovery, overall, is that rocky extrasolar planets have bulk elemental compositions similar to those of Earth and other rocky objects in our solar system. The white dwarf studies reveal that many extrasolar minor planets (asteroids) are differentiated, possessing analogs of terrestrial crust, mantle and core; this finding has important implications for the origin of our own solar system.

## 1) Overview

Discovery of planetary systems around stars other than our Sun has been a dream of scientists extending back centuries in time. This fascination derives, no doubt, from the question "Are we alone in the Universe?"

Well before the first extrasolar planets were discovered astronomers had considered a variety of planet detection techniques. These are discussed in other chapters in this book and include the indirect techniques of astrometry, transits, and precision radial velocity. In addition, one can directly image extrasolar planets via reflected starlight at visible wavelengths or by infrared radiation emitted by self-luminous warm planets. So-called "microlensing" is an additional indirect planet discovery technique that was first noted theoretically only as recently as 1964 (Liebes 1964). But it then took 40 years for this technique to yield its first planet (Bond et al 2004).

The purpose of the present chapter is to describe a planetary system discovery technique that is applicable only to white dwarf stars. Whereas all of the above techniques were realized theoretically well before they were observationally successful, the technique of identifying planetary systems around white dwarfs came literally out of the blue -- absolutely no one thought of it until after it was already observationally successful. That is, the observations came first to be followed only much later by proper interpretation of the data.

The white dwarf data -- primarily in the form of broadband and high-resolution spectra -- are unique in that they enable astronomers to measure the bulk elemental compositions of the extrasolar planetesimals that are the building blocks of rocky planets. Another unusual aspect of white dwarf planetary systems is that, typically, they orbit around stars that initially were more massive than the Sun, in contrast to the planetary systems detected via all of the above listed techniques that typically orbit stars of initial mass less than or about equal to the solar mass.

The first observational evidence of the existence of an extrasolar planetary system around a white dwarf, and indeed around any star other than the Sun, occurred 100 years ago when van Maanen (1917, 1919a, 1919b) discovered the closest single white dwarf to Earth, now named "van Maanen 2". But the proper interpretation of van Maanen's spectroscopic data -- that we now know indicate the existence of a planetary system surrounding this white dwarf (see Section 6A) -- came only 90 or so years later (Zuckerman 2015).

**2) Modern Picture of White Dwarf Planetary Systems**

A) Observational Basis

A typical white has a mass of about 0.6 solar masses, but a radius only about equal to that of Earth, and thus an enormously strong gravitational field. White dwarfs of interest in the present discussion have cooling ages in the range from 100 to 1000s of millions of years. Such stars have effective surface temperatures between about 20,000 K and 5000 K. Because of the strong gravitational field, elements heavier than helium should sink rapidly out of sight in times much shorter than these cooling ages (Schatzman 1945). Thus, the spectra of undisturbed white dwarf atmospheres should appear as hydrogen-rich (called DA white dwarfs), or helium-rich (DB) or as a mixture of these two elements. We now know, however, that about 30% of white dwarf atmospheres in the above temperature range display lines of one or more elements heavier than He (Zuckerman et al 2003, 2010; Koester et al 2014). Astronomers often refer to such white dwarfs as "polluted".

Where do these heavy elements come from and how are they maintained in the atmospheres of so many white dwarfs? One possibility is that elements are mixed up from the stellar interiors. The progenitors of many white dwarfs synthesized the element carbon in their interiors, and some white dwarfs (DQ-type) dredge this carbon up into their atmospheres where it is visible from Earth. However, the overall characteristics of the spectra of most polluted, non-DQ, white dwarfs are such that no plausible dredge-up model has ever been proposed. Rather the consensus has been that the heavy elements have been recently accreted onto the stars from some outside source. The reason that the accretion event must be recent is because typical heavy element settling times out of the stellar atmospheres range from days to millions of years -- all short time scales compared to the white dwarf cooling times (Koester 2009). Qualitatively, the settling times are shorter the warmer the temperature of the star. Also, at any given temperature, settling times for DA stars are much shorter than for DB stars.

For decades a popular accretion model invoked the interstellar medium (ISM) as the source of the observed heavy elements. The idea was that as a white dwarf moved through a relatively dense portion of the ISM the star's strong gravitational field would gather in the ISM gas and dust. A potential stumbling block to this model is that the ISM that surrounds the Sun is of low density. However, because the earliest known polluted white dwarfs were cool and possessed helium dominated atmospheres -- and thus relatively long heavy element settling times -- it was suggested that the observed heavy elements were accreted long ago when the polluted white dwarfs passed through a dense interstellar cloud quite far from the Sun.

This model took a direct hit when the first polluted DA star, G74-7, was discovered (Lacombe et al 1983); the heavy element settling time is only about 10,000 years -- too brief for the star to have traveled far in the ISM. So a new model, involving accretion by G74-7 of a comet, was proposed (Alcock et al 1986). Then, in 1987, infrared (IR) emission in excess of the photospheric emission was discovered from G29-38, a nearby DA white dwarf (Zuckerman & Becklin 1987). Graham et al (1990) demonstrated that the source of the IR excess is a dust cloud in close orbit around the white dwarf and suggested that the dust is due to the breakup of an asteroid that previously orbited the white dwarf. Soon thereafter the photosphere of G29-38, like G74-7, was found to be polluted with heavy elements (Koester et al 1997), suggesting a possible connection between the dust cloud and the presence of heavy elements in the photosphere.

The next major step forward was taken by Debes & Sigurdsson (2002) who pointed out that mass loss by the progenitor of a white dwarf could cause instabilities in a surrounding planetary system that could result in drastic changes in the orbits of planets and smaller objects, including injection in toward the central star. The beginning of the end of the ISM accretion model occurred in 2003 when a survey of ~100 cool DA stars revealed that ~25% contain calcium in their spectra (Zuckerman et al 2003). (For white dwarfs with temperatures in the range between the coolest white dwarfs and 20,000 K, calcium is usually the easiest element to detect optically -- at the highest temperatures magnesium becomes as easy to detect). In this 2003 paper, Zuckerman et al argued that none of the models that involve the ISM, comets, or mass transfer from a companion (seen or unseen), could explain their data set and that only an asteroid accretion model was not in conflict with the data. At about the same time Jura (2003) proposed a quantitative asteroid destruction model to reproduce the excess IR emission at G29-38. Figure 1 is an artist's conception of a white dwarf surrounded by a rocky asteroid belt far from the star and a hot dusty disk close to the star; the disk is the broken up remnant of one or more former asteroids.

In 2003 there were perhaps 50 known polluted white dwarfs -- about equally divided among stars with hydrogen and helium dominated atmospheres -- but only one star, G29-38, with excess IR emission indicative of a demolished asteroid. It took from 1987 to 2005 for discovery of an additional star with an IR excess (Kilic et al 2005, 2006; Becklin et al 2005). But, thanks primarily to the Spitzer Space Telescope, we now know that a few % of white dwarfs possess detectable, hot, dusty disks (Farihi et al 2009; Barber et al 2012). Thus about one in ten polluted white dwarfs have detectable disks; these white dwarfs tend to be among the most heavily polluted. A handful of these dusty disks also have detectable gaseous components (e.g., Gaensicke et al 2006; Jura 2008; Melis et al. 2010; Farihi 2016).

Our current observational picture of white dwarf planetary systems is based on ultraviolet, optical, and infrared spectroscopy and infrared photometry. The data are of the white dwarf photospheres and emission from surrounding hot dust and gas. Attempts have been made to directly image in the infrared warm, large, planet-mass companions to white dwarfs (e.g., Mullally et al. 2007; Farihi et al. 2008; Kilic et al 2009). To date the only known planet-mass companion was imaged with the IRAC camera on the Spitzer Space Telescope (Section 6F; Luhman et 2011) at the large projected separation of 2500 AU.

Various observational patterns among polluted and dusty white dwarfs have supported a model of accretion of rocky asteroids and ruled out all competing models. These patterns include, for example, carbon deficiency (Jura 2006), Galactic location (Farihi et al 2010), 10 μm silicate emission from circumstellar dust (Jura et al 2009), and mass budgets (Klein et al 2010, 2011). Carbon deficiency is characteristic of rocky objects in our solar system, and not of interstellar medium gas and dust. Figure 2 illustrates the low carbon abundance characteristic of all known polluted white dwarfs with the exception of one (WD 1425+540, see Section 6K). Galactic location demonstrates lack of correlation with dense regions of the ISM (from which heavy elements might be accreted). The minimum accreted masses in heavily polluted white dwarfs are far in excess of that of any known comet and the relative lack of volatile elements, such as carbon and nitrogen, is also inconsistent with a comet accretion model. With one notable exception at the time of this writing (Xu et al 2017), the lack of volatiles also rules out Kuiper Belt analog objects as the dominant source of accreted material.

Probably the most dramatic confirmation of the violent asteroid destruction/accretion model for polluted white dwarfs has come from observation and analysis of WD1145+017. Here an asteroid is being torn apart as we watch (Section 6J; chapter by Vanderburg and Rappaport in "Handbook of Exoplanets" 2018).

B) Theoretical Considerations

During the asymptotic giant branch (AGB) phase of stellar evolution that precedes the white dwarf stage, rapid mass loss and high stellar luminosities will greatly impact surrounding planetary systems. The F- and A-type main-sequence progenitors of most white dwarfs will typically expand to radii of a few AU or greater thus destroying all interior planetary objects. Because the mass loss is slow compared to orbital periods, surviving planets will spiral outward, increasing their semi-major axes by factors of two or more (a factor of 3 would be typical), i.e., the ratio of the initial stellar mass to that of the resulting white dwarf. (As noted in Section 2A, typical white dwarf masses are 0.6 times that of the Sun.)

Orbiting objects that are not engulfed by the AGB star will tend to spiral outward gently and in lock step. Nonetheless, as first emphasized by Debes & Sigurdsson (2002), orbital instabilities can arise. For example, due to the reduced mass of the central star, planetary Hill spheres will increase in radius. Then previously stable orbits of some small bodies located near a major planet -- for example, the Sun's asteroid belt and Jupiter -- will become unstable. Some altered orbits can intersect the tidal radius of the white dwarf resulting in asteroid break up and, due to loss of energy, an eventual settling into a close orbit around the white dwarf, as observed (Veras et al 2014a). (A white dwarf tidal radius is typically about the radius of our Sun.)

Altered orbits can also be a consequence of the YORP effect during the AGB evolutionary phase (Veras et al 2014b), or collisions between objects at any time, the debris of which goes off in various directions. Some of these pieces may have orbits that intersect the tidal radius of the white dwarf. Analysis of the composition of material accreted onto the atmospheres of some white dwarfs suggests that such collisions have indeed occurred (see Sections 4 & 6).

For white dwarfs in binary systems (e.g., Zuckerman 2014), the Eccentric Kozai-Lidov mechanism (Naoz 2016; Stephan et al 2017) can alter orbits of objects even as large as planets, such that the resulting orbit intersects the tidal radius of the white dwarf. It is interesting to speculate about the observational characteristics of a system where a white dwarf accretes an entire planet.

Because white dwarfs do not possess particle winds and because radiation pressure is small for the temperature range of interest here (<20,000 K), once material is in close orbit around a white dwarf it will not be ejected from the system. Eventually the orbiting material must be accreted onto the white dwarf. A minimum accretion rate is set by Poynting-Robertson (radiation) drag (Rafikov 2011a), but sometimes the accretion rate is much greater, probably driven by viscosity in the disk gas (Rafikov 2011b, Metzger et al 2012, Jura 2008).

A luminous AGB star can vaporize volatile components (for example water ice) of orbiting objects. However, calculations (Jura & Xu 2010; Malamud & Perets 2016, 2017) indicate that, for objects that are sufficiently massive or far from the central star or both, most of the volatile material will be retained. Thus, the apparent paucity of water ice in most parent bodies that have been accreted by polluted white dwarfs is probably due to an initial dry formation process.

Although neither an asteroid belt nor major planets have ever been directly imaged at white dwarf stars (with one unusual planet-mass exception, Section 6F), the totality of the weight of the evidence mentioned above and below is so compelling that there can be little doubt that such objects commonly orbit white dwarfs and their progenitors. These planetary system components are present at white dwarfs out beyond the regions best probed by the transit and precision radial velocity techniques used to study planets around main sequence stars.

Accretion of rocky material commonly occurs onto white dwarfs with cooling age as long as 1000, or more, million years. This has been modeled to indicate the existence of tightly packed multiple major planets at large semi-major axes, perhaps as many as 4 or more such planets (Veras & Gaensicke 2015). Such systems would resemble our own solar system with its 4 tightly packed major planets (by "tightly packed" we mean that there is not room in semi-major axis space to fit in another planet). Another example is the A-type star HR 8799, again with 4 tightly packed major planets (Marois et al 2010). Such tightly packed, but extensive, systems could be regarded as the extended counterparts of the tightly packed close-in planets discovered by the Kepler Space Telescope. Thus, a combination of the Kepler and white dwarf studies implies that many stars host a variety of planets with orbits that encompass a wide range of semi-major axes, much as in our own solar system.

Recent major reviews related to post-main-sequence planetary system evolution include Jura & Young (2014), Farihi (2016) and Veras (2016)

**3) Sampling the Compositions of Small Bodies with White Dwarfs**

The standard model for the sampling of rocky and/or icy bodies by white dwarfs is that perturbations of the orbits of these bodies lead to passage within the tidal radius of the white dwarf. As a result, the bodies can be shredded to form a debris disk around the central star (Debes & Sigurdsson 2002). Under these violent conditions, it is likely that portions of planetesimals or minor planets impact the white dwarf.

We use ratios of elements to compare polluted white dwarf compositions with solar-system bodies and their various components (core, mantle, crust) in order to mitigate the uncertainties surrounding the nature of the accretion process. The advantages of using element ratios can be seen using a simple equation that characterizes the time-dependent mass of an element $Z$ in a polluted white dwarf convective layer or photosphere:

$$\frac{dM_z}{dt} = \dot{M}_z - \frac{M_z}{\tau_z}$$

where $M_z$ is the mass of element $Z$ in the white dwarf surface regions, $\dot{M}_z$ is the accretion rate of the element onto the star, and $\tau_z$ is the characteristic time for settling out of the convective layer or photosphere for element $Z$. This first-order linear differential equation has the usual general solution

$$M_z = ce^{-t/\tau_z} + e^{-t/\tau_z}\int e^{t/\tau_z}\dot{M}_z(t)\,dt$$

where *c* is an integration constant that is zero where the mass of Z at time zero is also zero (the typical assumption). For the He-rich DB white dwarfs, the settling times are expected to be long due to the need to settle through large convective layers. In these cases the large values for $\tau_z$ suggest that often $t/\tau_z \ll 1$ so that $\exp(-t/\tau_z)$ and $\exp(t/\tau_z)$ approach unity and the solution becomes

$$M_z = \int \dot{M}_z(t)\,dt .$$

We can assume that during this buildup of element Z in the convective layer the rate of accretion is effectively constant, yielding

$$M_z = \dot{M}_z t ,$$

where *t* is the timescale over which the element is being added. For any two elements, Z1 and Z2, taking their ratio eliminates the time variable such that the mass ratio in the white dwarf faithfully reflects the same ratio in the parent body that is being accreted:

$$\frac{M_{z2}}{M_{z1}} = \frac{\dot{M}_{z2}}{\dot{M}_{z1}} .$$

This shows clearly the advantage of using elemental ratios; timescales cancel and measured ratios can be interpreted directly as the composition of the infalling material.

In the hydrogen-rich DA white dwarfs, the timescale for settling will be much shorter. In these cases $t/\tau_z > 1$ such that the product $\exp(-t/\tau_z)\exp(t/\tau_z)$ approaches unity and $\exp(-t/\tau_z)$ approaches zero, so the solution for $M_z$ with constant accretion rate becomes

$$\begin{aligned}M_z &= e^{-t/\tau_z}\int e^{t/\tau_z}\dot{M}_z(t)\,dt \\ &= e^{-t/\tau_z}\left(\dot{M}_z \tau_z e^{t/\tau_z} - \dot{M}_z \tau_z\right) \\ &= \dot{M}_z \tau_z\end{aligned}$$

and the ratio of the accreted mass of two elements in this case is

$$\frac{M_{z2}}{M_{z1}} = \frac{\dot{M}_{z2}\tau_{z2}}{\dot{M}_{z1}\tau_{z1}} .$$

In these circumstances it is therefore necessary to estimate the settling times of the different elements in order to use the white dwarf pollution as a measure of the element ratios in the impacting bodies. Settling times depend on white dwarf temperature and gravity, quantities that can be modeled. Fortunately, settling times for most rock-forming elements under a wide range of conditions are similar to one another within approximately ten per cent (Koester 2009). Exceptions are at the extreme end of the mass range of detected elements, where, for example, lighter elements like carbon can be twice as slow to settle as other elements in DA stars with relatively low temperatures, or where (heavy) iron can settle twice as fast under certain conditions. This means that, in the worst cases, element ratios can be uncertain by as much as a factor of two where settling times are poorly constrained by modeling.

Other circumstances may also occur, including non-steady-state conditions where settling continues after accretion ends or where accretion is time variable. More complex solutions to the mass-balance equations that represent these cases have been investigated by Jura and Xu (2012).

The above equations show that maximum masses of the impactors can be estimated from the cooling ages of the white dwarfs by assuming that the pollution continued throughout the lifetime of the white dwarf as a limiting case. A typical average accretion rate for a population of DB white dwarfs of $1.4 \times 10^8$ g/s (Jura and Xu 2012; Gänsicke et al. 2012), combined with their mean cooling age of 230 Myr, yields an average maximum mass of $1 \times 10^{24}$ g of accreted material, or roughly the mass of 1 Ceres the largest asteroid in our solar system. Typical dusty disk lifetimes are thought to be much shorter than the cooling ages of the white dwarfs, closer to $10^5$ to $10^6$ years, meaning that many bodies accreted by the white dwarfs are smaller than 1 Ceres and more like the mass of asteroid 21 Lutetia ($\sim 10^{21}$ g) in the asteroid belt.

**4) Comparisons with the Bulk Compositions of Solar System Bodies**

The rock and ice-forming elements found in polluted white dwarfs allow for the examination of the geochemistry of small bodies that orbit other stars. In the solar system, the four elements O, Mg, Si, and Fe comprise more than 90% of the mass of rocky bodies. Thus far the white dwarf data show no significant departures from this pattern, indicating that rocky bodies around other stars are similar to those in the solar system.

Because of the striking similarities between variations in the ratios of rock-forming elements among different polluted white dwarfs and those produced by igneous processes (e.g., Gaensicke et al 2012), Jura and Young (2014) suggested a new field referred to as "extrasolar cosmochemistry" (cosmochemistry is commonly defined as the study of the chemistry of meteorites and the rocky bodies from which they derive in the solar system). The range in polluted white dwarf ratios of the rock-forming elements shows that the impacting bodies include metal-rich cores, rocky mantles, and highly evolved crust that are indistinguishable from cores, mantles, and basaltic crusts in our own solar system. For example, large variations in Fe/Si together with limited ranges in Mg/Si are consistent with igneous differentiation of the impactor bodies into metal cores and rocky mantles. Similarly, ratios of Si/Al and Mg/Al covary among the white dwarfs in a manner that mimics igneous differentiation into basaltic crust and mantle (Figure 3). The variations of element ratios seen in the white dwarfs are centered on the element ratios of chondritic meteorites, solar abundances, and F- and G-type main sequence stars that together define a reference bulk composition (Figures 3 and 4). This observation suggests that the extrasolar rocky bodies that impacted the white dwarfs were indistinguishable from chondrites in bulk composition.

The specificity of extrasolar cosmochemistry afforded by the polluted white dwarfs is limited by the precision of the measurements. However, in some cases, the conclusions can be strikingly detailed. For example, Xu et al. (2013) showed that the body accreted by white dwarf GD 362 is similar to a class of meteorites called mesosiderites. Mesosiderites are mixtures of metal and basaltic crust that are thought be to the products of collisions between plantesimals. This mixing is evidenced in the element ratios of GD 362, suggesting that collisions between small bodies and their products are present in extrasolar planetary systems as well.

The volatile concentrations for the impacting objects can be obtained by direct observation or, in the case of oxygen that was bound in water molecules in the parent body, by a difference technique (see following paragraph). Rocky planetesimals and planets in the solar system, as represented by Earth and the asteroids sampled by meteorites, are depleted in carbon by at least a factor of 10 relative to the Sun (e.g., Figure 2, Lodders 2003, Jura 2006). The white dwarf data show that the same is usually true of extrasolar planetsimals and minor planets (with few exceptions, Figure 2; Wilson et al. 2016). The reasons for the paucity of carbon in rocky bodies is debated, but the white dwarf data establish that whatever this process is, it is fundamental to rock formation around stars.

Because water dissociates to oxygen and hydrogen upon being engulfed by a white dwarf, water contents are derived by difference. This is done by assigning each rock-forming element (and carbon if present) an amount of oxygen required by charge balance (e.g., $SiO_2$, $MgO$, $CaO$, $Al_2O_3$, $CO$, etc.) then subtracting these molar abundances of oxygen from the total measured oxygen content of the white dwarf. The difference is assumed to represent water. While most accreted objects are poor in $H_2O$ and volatile elements (Jura & Xu 2012; Gentile Fusillo et al 2017), a few exceptions are described in Section 6. A key advantage afforded by the white dwarf data is that they can represent samplings of large fractions of the impactors. This is the only way, for example, that we are likely to obtain a measurement of the bulk elemental composition of a Kuiper-belt like object, as has now been done for one polluted white dwarf (see Section 6K).

**5) Implications of Extrasolar Planetesimal Differentiation for the Origin of the Solar System**

The existence of cores, crusts and mantles among the polluting rocks that have impacted the white dwarfs, combined with their relatively small masses -- comparable to those of the asteroids in the Sun's asteroid belt, e.g., 4 Vesta ($2.6 \times 10^{23}$ g), 1 Ceres, and smaller asteroids -- has important implications for the history of our solar system.

The heat to melt and therefore differentiate asteroid-sized bodies into cores, mantles and crusts comes primarily from the decay of the short-lived radioisotope $^{26}$Al with a half-life of 0.7 million years. Other heating sources, including conversion of gravitational potential energy upon accretion and frictional heating due to collisions are ineffective at these masses. The origin of $^{26}$Al in the solar system has long been debated. For many years the existence of this and other short-lived radionuclides were attributed to close encounters with a supernova or another particular exotic stellar source. In fact, excesses of the short-lived radionuclides were touted as evidence for a supernova trigger for the formation of the solar system (Cameron & Truran 1977). The fact that melting of asteroid-sized bodies occurred in planetary systems around many of the polluted white dwarfs shows that $^{26}$Al was extant in these systems as well. This observation led to a reassessment of the meaning of $^{26}$Al in our own solar system – this showed that the concentration of this nuclide in the solar system was in fact normal for star-forming regions in general (Jura et al 2013, Young 2014, Young 2016). The ubiquity of $^{26}$Al in star-forming regions in turn has led to the suggestion that all of the short-lived radionuclides evidenced in meteorites reflect the formation of the solar system in a large star-forming region like those observed today in our Milky Way Galaxy. An important source of these nuclides appears to be winds from Wolf-Rayet stars (Young 2016).

**6) Individual Polluted White Dwarfs of Special Interest**

In most polluted white dwarfs observed optically, only calcium is detected at both high (Zuckerman et al 2003) and low (Sloan Digital Sky Survey [SDSS] data releases) spectral resolution, while, in the ultraviolet, usually only silicon has been seen (Koester et al 2014). At this point in time a few hundred white dwarfs polluted with heavy elements are known, but this number should increase greatly as a result of future all-sky surveys similar to the SDSS. When many elements are detected in an ensemble of white dwarfs, overall commonalities and patterns can be recognized as discussed in Sections 3 and 4. In the present section we note some individual stars that stand out from the crowd by virtue of their special place in history or their unusual characteristics.

A) van Maanen 2: the first known polluted white dwarf

Before astronomers even understood what a white dwarf star was, van Maanen (1917 & 1919a) detected Fe, Ca and Mg in what we now know to be the closest single white dwarf to Earth, "van Maanen 2". (Two white dwarfs in binary systems are closer to Earth than is van Maanen 2.) As noted in Section 1, we now understand that the presence of these elements in van Maanen 2 was the first observational evidence -- by any technique -- of the existence of an extrasolar planetary system. van Maanen 2 is a helium atmosphere (DB) white dwarf as were all subsequently detected polluted white dwarfs until G74-7.

B) G74-7: the first known polluted, hydrogen-atmosphere (DA), white dwarf

Lacombe et al (1983) detected Ca in G74-7. Because of the relatively short residence time of the Ca in the photosphere of G74-7 (~$10^4$ years), Alcock et al (1986) proposed a model of accretion of a comet by the star. They also noted that the absence of Ca in the atmosphere of other DA stars could be used to set limits to the properties of Oort-like comet clouds around such stars.

C) G29-38: the first white dwarf known to possess a hot dusty orbiting disk

While searching for brown dwarfs around white dwarfs, Zuckerman & Becklin (1987) detected excess infrared emission from G29-38. A few years later Graham et al (1990) showed that the IR excess is due to an orbiting disk of dust particles and they suggested that the material in the disk might be due to a busted up asteroid. Jura (2003) fully developed this model in such a way that quantitative properties of future detected disks could be deduced. It took 18 years for a second dust disk to be identified (around GD 362; Kilic et al 2005; Becklin et al 2005).

D) SDSS J1228+1040: the first white dwarf known to possess an orbiting gaseous disk

While searching through white dwarfs in the Sloan Digital Sky Survey, Gaensicke et al (2006) recognized lines of Ca and Fe in emission from the single white dwarf SDSS J1228+1040 and realized that these lines must be generated in a hot orbiting gaseous disk. Later Melis et al (2010) argued that the lines were coming from what they called a "ZII-region" where the dominant heating and cooling processes involve heavy elements, rather than hydrogen as in conventional HII-regions well known throughout astronomy. Melis et al also showed that the dust and gaseous disks around a few white dwarfs were more or less spatially coincident, and within the tidal radii of the white dwarfs. Manser et al (2016) show that the gas disk is precessing with a period of ~25 years possibly due to effects of General Relativity.

E) GD 362: a white dwarf with 17 identified elements

To date, the king of the polluted white dwarfs is GD 362 with 15 heavy elements plus H and He seen in its photosphere (Zuckerman et al 2007). These elements include strontium (Sr) and scandium (Sc) that, in the Sun, are down in abundance by a factor of a billion compared to the dominant element hydrogen. Detection of such rare elements in white dwarf spectra illustrates this technique's incredible sensitivity to elemental composition of planetary material when compared with all other methods of study of extrasolar planetary systems.

F) WD 0806-661: a white dwarf with a distant, planet-mass, companion

Although the only plausible model for heavy-element pollution of (most) white dwarf atmospheres implies the presence of one or more orbiting major planets, WD 0806-661 is the only white dwarf where a planet-mass object has actually been seen directly (Luhman et al 2011). When the white dwarf was a main sequence star, the semi-major axis of the companion was at least 800 AU. If the star has an age of 1.5 Gyr since birth, then the companion has an estimated mass about seven times that of Jupiter.

G) GALEX J193156.8 + 011745: a heavily polluted white dwarf first noticed via its excess UV emission

This bright white dwarf was first recognized as UV bright and heavily polluted in an ultraviolet/optical survey (Vennes et al 2010). As a consequence of its brightness, strong IR excess emission, and abundance of heavy elements, the star has been carefully studied (Vennes et al 2011, Melis et al 2011, Debes et al 2011).

H) GD 61: evidence for a wet planetesimal

As noted in Section 4, most planetesimals accreted by white dwarf stars appear to be dry, that is, all of the measured oxygen can be accounted for by accretion of oxides of the abundant elements Fe, Mg, SI, and Ca without any need to include a water component. GD 61 is a notable exception; according to the analysis by Farihi et al (2013), the accreted planetesimal is especially rich in oxygen. The excess oxygen can be accounted for if the parent body was originally composed of 26% water by mass.

I) NLTT 43806: evidence for an extrasolar planetary lithosphere?

The heavy element pollution in this unusual white dwarf was first noted and studied by Kawka & Vennes (2006). Zuckerman et al (2011) found that the body accreted onto NLTT 43806 was aluminum and calcium-rich, and iron-poor. They compared the relative abundances of Al and eight other heavy elements seen in NLTT 43806 with the elemental composition of bulk Earth, with simulated extrasolar rocky planets, with solar system meteorites, with the atmospheric compositions of other polluted white dwarfs, and with the outer layers of the Moon and Earth. The best agreement was found with a model that involves accretion of a mixture of terrestrial crust and upper mantle material onto NLTT 43806. Jura et al (2014) proposed that the barium to calcium abundance ratio in material accreted onto white dwarfs might be used to distinguish between continental and oceanic crust in the parent body. (Barium has not yet been detected in any polluted white dwarf.)

J) WD 1145+017: observations of an asteroid being torn apart in real time

WD1145+017 was contained in a K2 field of view of NASA's Kepler space telescope. Vanderberg et al (2015) noted dips in the optical brightness of the white dwarf with periods ranging from 4.5 to 4.9 hours. They attributed these dips to the passage between Earth and the star of several chunks of material from a broken-up planetesimal. Independently, Xu et al (2016) identified 11 heavy elements in absorption either in the photosphere or a gaseous orbiting disk or both. The gaseous material has the wide velocity dispersion of ~300 km/s that is difficult to explain with any simple kinematic model. Both the broadband and the spectral line absorptions change noticeably with time (Figure 5). Detailed discussions of this star can be found in Veras et al (2017) and the chapter by Vanderburg and Rappaport in "Handbook of Exoplanets" (2018).

K) WD 1425+540: a planetesimal rich in nitrogen; evidence for a Kuiper Belt analog

object

As noted above, most planetesimals accreted onto white dwarfs are dry with the occasional wet exception (e.g., GD 61). The accreted objects at some white dwarfs also contain a modest quantity of carbon (Koester et al 2014; Wilson et al 2016); in all these cases the objects could come from an asteroid belt analog. In contrast, Xu et al (2017) identified a white dwarf with an optical/ultraviolet spectrum that implies substantial quantities of water, carbon, and, of most interest, nitrogen in the accreted parent body. This is the first white dwarf with N reported in its spectrum. The presence of so much N implies that the parent body formed in a cold region far from the main sequence progenitor of the white dwarf, a region analogous to the Sun's Kuiper Belt. Such volatile-rich objects might have supplied much of Earth's water. WD 1425+540 is orbited by a distant companion star; its gravity might have perturbed the Kuiper Belt analog object in towards the white dwarf via the Eccentric Kozai-Lidov mechanism. (Stephan et al 2017).

L) SDSS J1043+0855: evidence for a differentiated, rocky, carbonate-containing body?

Melis & Dufour (2017) describe material accreted onto this white dwarf as rocky and iron-poor, but containing substantial quantities of carbon. They consider the accreted material to come from the outer layers of a differentiated object similar perhaps to NLTT 43806 (bullet I above), but noting that, for SDSS J1043+0855, the carbon might indicate the presence of calcium carbonates in the accreted parent body.

**Dedication:** UCLA Professor Michael Jura was a pioneer in the use of polluted white dwarfs as uniquely remarkable probes of extrasolar planetary systems; his creative ideas and contributions enlightened the field.

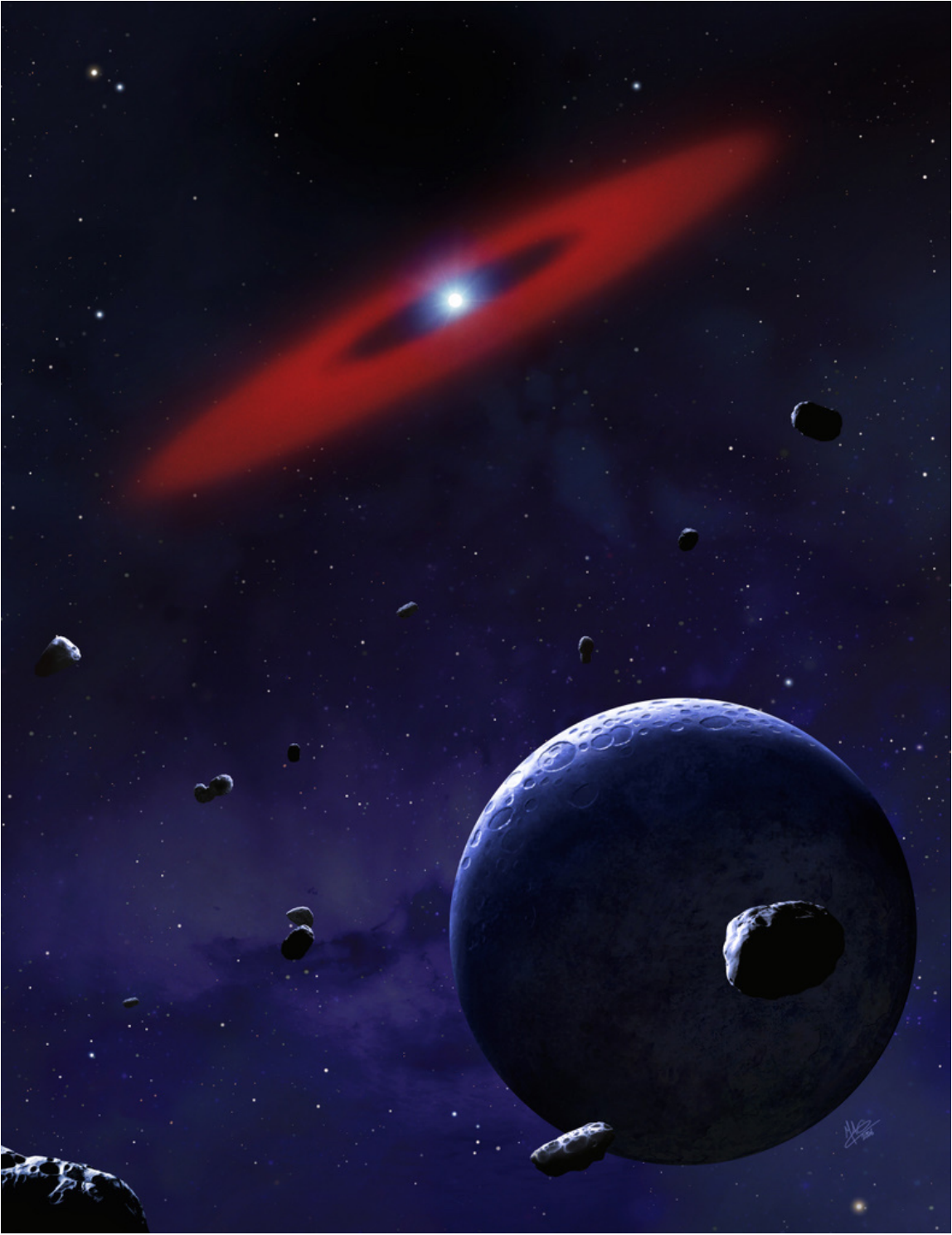

Figure 1: Artist's conception of a white dwarf surrounded by asteroids and a hot dusty disk composed of material from one or more former asteroids. The hot dust orbits within a region with a diameter similar to that of our Sun while the white dwarf itself is about the size of Earth. (Credit: Mark A. Garlick / space-art.co.uk)

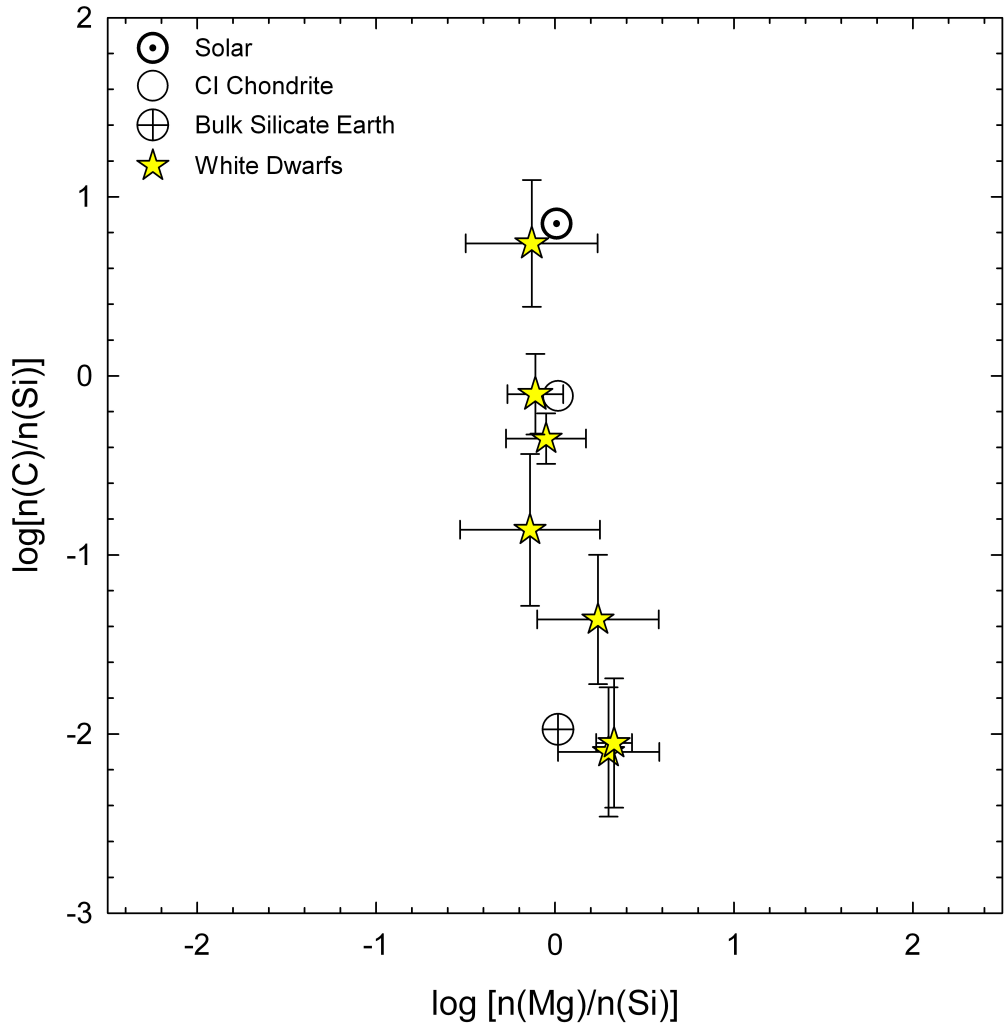

Figure 2: Plot of C/Si vs Mg/Si by number of atoms illustrating the range in carbon abundances measured in polluted white dwarfs. The highest relative carbon concentration is for WD 1425+540 (the yellow star next to the symbol for the Sun; see Section 6, item I) that has an essentially solar value and has been polluted by a Kuiper Belt analog object rather than an asteroid-like object (Xu et al 2017). The second highest carbon value is for Ton 345, with C/Si data from Wilson et al. (2015) and Mg/Si data from Jura et al. (2015). Note that, in contrast to the widely variable C/Si ratios, the magnesium to silicon ratio is essentially constant in extrasolar planetesimals and is indistinguishable from CI chondrites, solar, and bulk Earth.

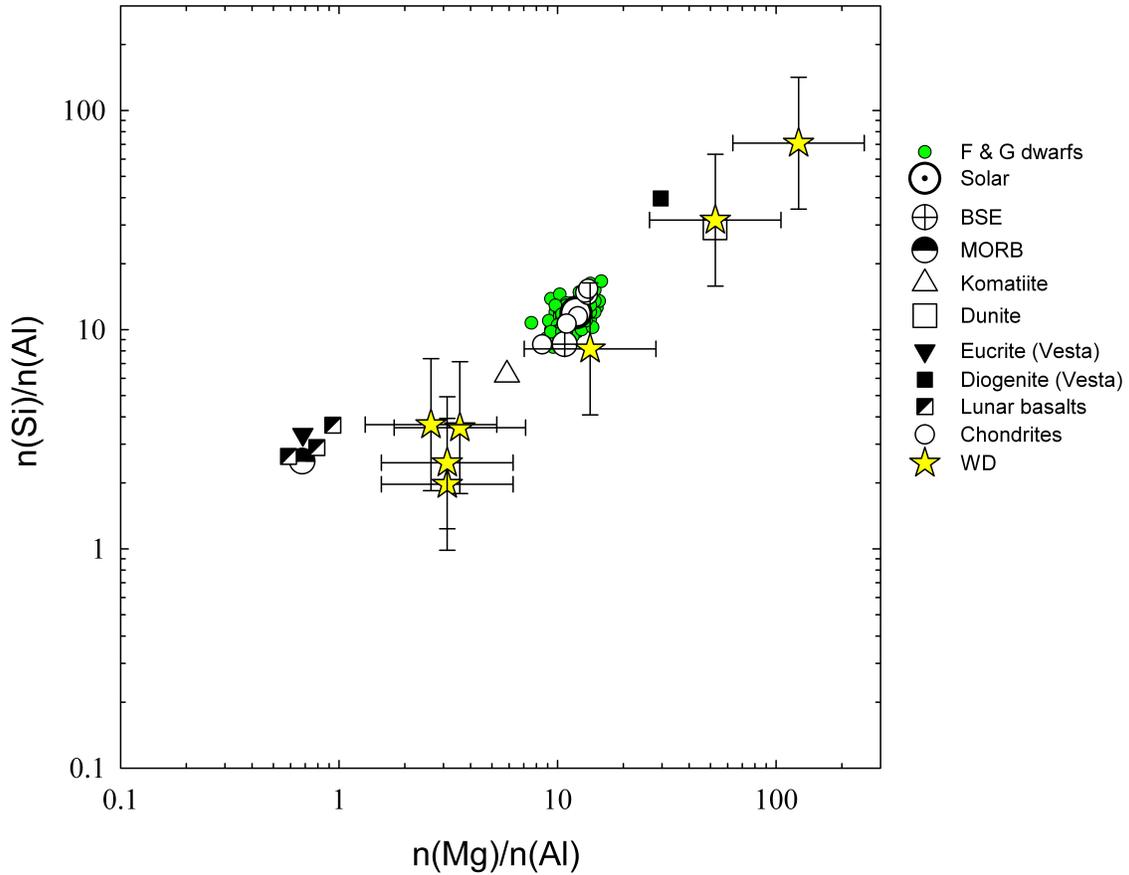

Figure 3: Si/Al vs Mg/Al ratios by number of atoms for polluted white dwarfs (yellow stars), F- and G-type stars (green circles), and various geological materials for reference. Also displayed are the bulk silicate Earth (BSE), solar, and CI chondrite meteorites (CI). The F and G star abundance ratios are similar to solar and chondritic (the solar symbol is hidden behind the chondrite data points). Diogenites and dunites (mostly olivine) represent residues after partial melting to form basalts that comprise most of the igneous (solidified) lava found on planetary surfaces, including terrestrial mid-ocean ridge basalts (MORB). Komatiite is a primitive type of mantle-derived igneous rock. The spread in white dwarf data is similar to the trends defined by igneous differentiation of mantle and crust. Data are from Jura and Young (2014) and references therein.

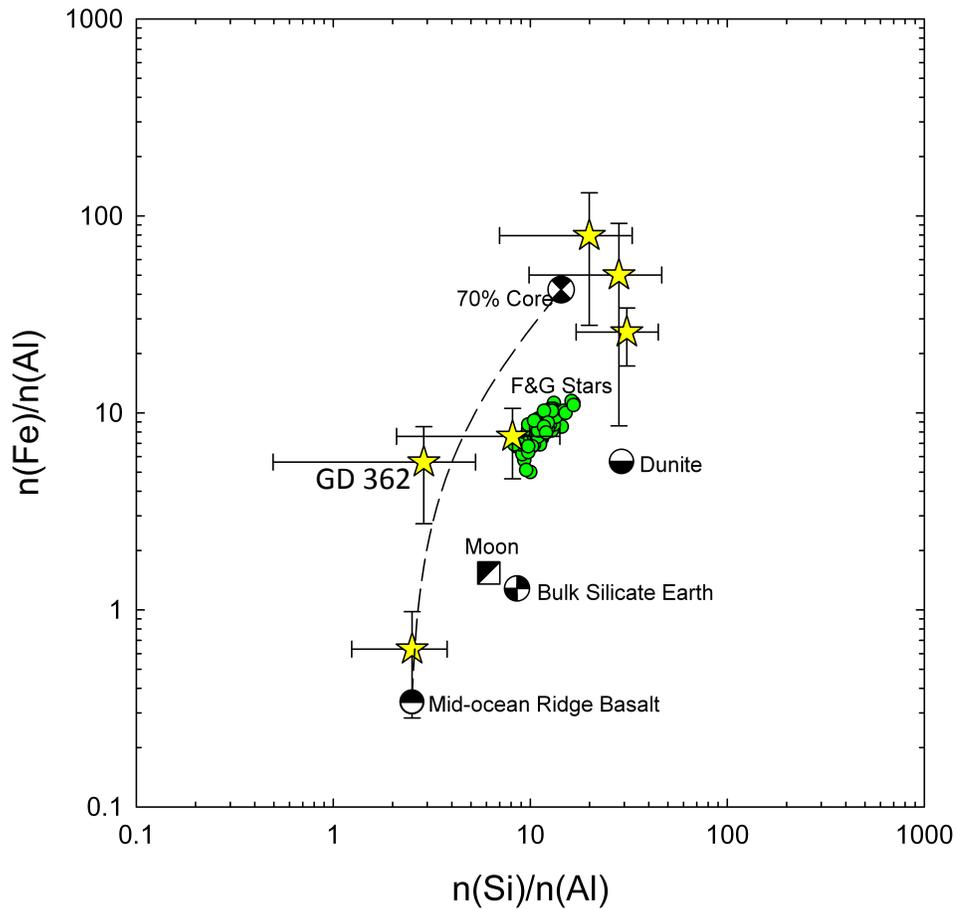

Figure 4: Fe/Al vs Si/Al by number of atoms for polluted white dwarfs (yellow stars) compared with several rocky body reference compositions, including bulk silicate Earth, terrestrial mid-ocean ridge basalts and dunites, and a calculated composition representing 70% core, 30% mantle by mole. F and G stars are shown for reference as well. A mixing curve (dashed line) between a metal core-rich composition and basaltic crust is shown for reference. The datum for white dwarf GD 362 is shown for comparison with the crust-core mixing curve. This object is proposed to be similar to mesosiderite meteorites that are also mixtures of metal and basalt (see Section 4). The lowest plotted yellow star is NLTT43806 (see (I) in Section 6). Data are from Jura and Young (2014) and references therein.

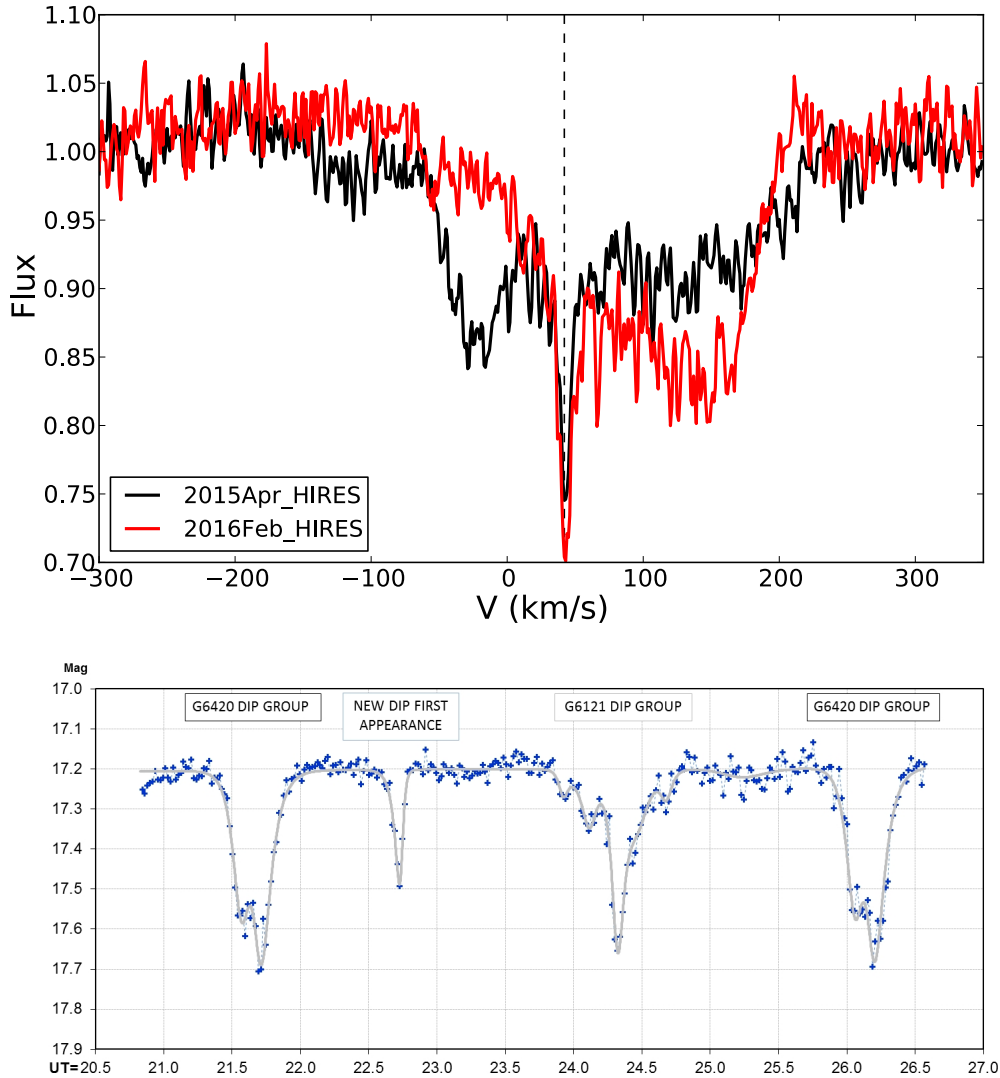

Figure 5: Secular variations in WD 1145+017. The upper panel is an average of several optical circumstellar absorption lines (Xu et al 2016 and private communication 2017). The narrow absorption feature indicated by the vertical dashed line is due to ions in the stellar photosphere; these lines do not noticeably vary with time. The broad absorption that extends from about -70 to +200 km/s is due to gaseous constituents of an orbiting circumstellar disk; as can be seen, the intensity of these lines varies strongly with time. The lower panel (from Gary et al 2017) displays absorption dips in the optical light curve of WD 1145+017 as pieces of a disrupted asteroid pass between the star and Earth. The abscissa is in hours and spans 1.3 orbits.